# Security Monitoring of the Cyber Space

Claude Fachkha
*Concordia University, Canada*

**ABSTRACT**

*Billions of users utilize the largest and more complex network of information, namely, the Internet. Despite the fact that this IT critical infrastructure provides various communication services, adversaries are abusing the Internet security and privacy to execute cyber attacks for various reasons. To cope with these threats, security operators utilize various security tools and techniques to monitor the cyber space. An efficient way to monitor and infer threat activities online is to collect information from trap-based monitoring sensors. This chapter primarily defines the cyberspace trap-based monitoring systems and their taxonomies. Moreover, it presents the state-of-the-art in terms of research contributions and techniques, tools and technologies. Furthermore, it identifies the gaps in terms of science and technology. Additionally, it presents some case studies and practical approaches corresponding to large-scale cyber monitoring systems such as Nicter. In this context, we further present some related security policies and legal issues on network monitoring. In a nutshell, the chapter aims to provide an overview on the Internet monitoring space and provides a guideline for readers to help them understand the concepts of observing, detecting and analyzing cyber attacks through network traps.*

## 1. INTRODUCTION

As of today, the Internet, the network of networks, provides information sharing and communication systems to more than 7 billion users [1]. This number is dramatically increasing as humans are becoming more dependent on social media/networks, mobiles, telecommunication, gaming, dating websites in addition to various cloud services and facilities. This increase rises the size of information sharing and hence created the term Big Data. This term has become the focus, the challenge and the exclamation mark for Internet Service Providers (ISPs), organizations, law enforcements and government agencies. For example, the questions are how to handle such large amount of information? How to analyze the traffic and how to secure and control such big data?

In regard to privacy, security and control, this cyberspace challenge takes an appeal of a continuous conflict due to the fact that computer attack tools and techniques are becoming more intelligently designed and hackers are capable of launching worldwide impacting attacks for various reasons such as large-scale Denial-of-Service [2], cyber-terrorism, information theft, hate crimes, defamation, bullying, identity theft and fraud. For instance, the Flame code, the most complex malware ever found, is a new generation of malware discovered in 2012 that aims to target nuclear power plants machines [3]. This obviously open the door of a new cyber war impacting the whole world including critical physical infrastructure such as power plants, nuclear, etc.. Further, the existence of widely available encryption and anonymizing techniques makes the surveillance and the investigation of cyber attacks much harder problem. In this context, the availability of relevant cyber threat collecting and monitoring systems is of paramount importance.

In this content, we attempt in this chapter to find answers to the following questions:

- How do we investigate large-scale cyber events?
- What is the state of the art of cyber security monitoring systems?
- Who has the capability to monitor the cyberspace?
- What are the privacy and security policies behind security analysis and deployment?

Answering the above questions can help security operators to understand the objectives of network monitoring technologies, their corresponding pros and cons, their deployments, their traffic analysis processes, their embedded threats, the difference among them as well as their research gaps. Moreover, the readers can have an overview on the security policies and legal issues that mostly unknown for network operators and even adversaries.

The rest of this chapter is organized as follows. Section II presents the background information on various network monitoring concepts and cyber threats. Section III provides the state of the art of trap-based monitoring tools. Section IV discusses the security policies and legal issues in network monitoring. Section V presents the trend and the future research directions. Finally, Section V summaries and concludes with a discussion on the monitoring systems and research.

## 2. BACKGROUND

This section provides an overview on the major elements of this chapter, namely, the trap-based monitoring systems and their embedded cyber threats.

### 2.1 Trap-based Cyber Security Monitoring Systems

The Internet involves several telecommunication elements and machines such as servers/clients, network infrastructure (routers, switches, etc.), services (email, web, etc.) and databases, among many others. One way to monitor the information sharing among these devices is to build network management tools based on SNMP services [4]. In addition to managing the Internet, several security monitoring systems exist online. These methods run over various network layers such as physical, network and application as well as covers various areas such as storage, access control, etc. Trap-based monitoring sensors are systems that aim to trap adversaries online. The purpose is to collect insights on the attack traces and activities such as probing/scanning for vulnerable services, worm propagation, malware downloads and other command-and-control activities such as executing DDoS cyber attacks using Botnet [5]. Figure 1 depicts a basic concept of trap-based sensor and monitoring on the Internet.

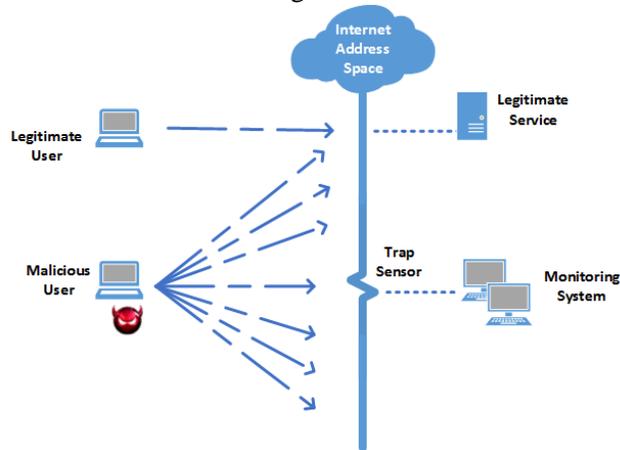

Figure 1: Basic Monitoring Systems using Trap Sensors

As shown in the figure above, a trap monitoring sensor is deployed on the Internet address space to attract malicious users. In some cases, these trap sensors run on unused but routable IP address, hence, all traffic destined to them is considered suspicious and therefore requires further investigation. The trap attracts adversaries by running vulnerable services. As shown in the above figure, once these attackers connect to the sensors, all the malicious traffic is forwarded from the sensor to the monitoring systems for further analysis. In the rest of this chapter, we discuss the state of the arts of these monitoring systems, their deployments, their embedded threats as well as their pros and cons.

## 2.2 Trap-based Embedded Cyber Threats

Several cyber threats exist on the Internet, we define below the major threats that can be found through the analysis of sensor-based monitoring systems.

**Scanning/Probing:** Scanning, also known as reconnaissance activities, is the first step in a cyber attack life cycle. Adversaries run scanning activities to infer vulnerabilities on the Internet. Once a machine is found vulnerable, the attacker attempts to control or infect this host based on the inferred vulnerability. For instance, an attacker check for vulnerabilities in remote access services first before trying to access remotely and take control. Scanning various in natures, strategies and approaches. Moreover, several scanning techniques exists such as stealthy, sweep, open and half open scans, among others [6].

**Botnet:** Botnet can be used as a platform for adversaries. It is designed to command and control compromised machines. Hence, it aims to provide a powerful tool for hackers distribute and amplify their attacks. A typical botnet consists of three major elements, a bot master, command-and-control (C&C) hosts and bots/zombies of compromised machines. A bot master controls an army of bots through C&Cs. For instance, a bot master can order all bots to flood (send extensively) large number of requests to one single victim to exhaust or/and deny its resources. Further, a bot master can send spams on the Internet through his botnet infrastructure. Several botnet infrastructure exists such centralized (IRC-based) and decentralized (P2P-based) [7].

**Denial of Service:** Denial of Service attacks are characterized by an explicit attempt to prevent the legitimate use of a service. DDoS attacks employ multiple attacking entities (i.e., compromised Machines/bots) to achieve their intended aim which mainly targets depleting bandwidth and host resources (e.g., CPU usage). In general, botnet is an impeccable tool to execute DDoS attacks with high impact [8].

**Exploit:** Exploit is a software or a sequence of commands that tackles bugs, glitches or/and vulnerabilities in computer system for the purpose of executing malicious acts such as gaining control, accessing root privileges, compromising and infecting exposed machines [9].

**Malware:** Malware is a piece of code designed to perform malicious activities. Malware comes with various types such as viruses, worms, trojans, etc. Each of these types has different features and families. Some of these features can include propagation and replication. For instance, worms are known to have self-propagating activities. Malware can damage computer services in many ways like in compromising machines, stealing information, taking control, corrupting system files, exhausting resources, etc. [10].

## 3. TRAP-BASED CYBER MONITORING SYSTEMS

- In this section, we provide a state-of-art study on the main trap-based monitoring systems. The purpose of this section is to offer the reader a guideline on monitoring systems, namely, darknet, IP gray space, honeypot, greynet, honeytokens, etc. Therefore, we aim to answer the following questions: How do we investigate large-scale cyber events? What is the state of the art of cyber security monitoring systems?

## 3.1 Darknet

Darknet, also known as network telescopes [11], Internet background radiation or blackhole, is a chunk of Internet address space running on unused but routable IP addresses. Therefore, all traffic destined to them is more likely suspicious. To the best of our knowledge, darknet research and analysis started in the beginning of the 1990th when Bellovin leveraged the unused address space to uncover malicious activities [12]. Few years later, darknet research and analysis started to become more attractive for researchers. At that time, darknet was found as a good source to collect security insights online. Some of the projects include the analysis of Internet threat through darknet such as DDoS [13], worms [14], scanning [15] as well as misconfiguration [16].

### 3.1.1 Darknet Deployment

This section represents an overview of darknet deployment. The first step in darknet technique is to deploy a sensor monitoring system. Therefore, understanding the network architecture is a most. Thus, a careful configuration must be done on the dynamic host server or the upstream router to forward unreachable packets to the darknet sensors. Two major elements must be done for this deployment setup, namely, the storage and network requirements and the deployment techniques. First, it is critical to identify the exact storage and network requirements for a darknet system. In order to collect darknet data, *PCAP* and *Netflow* formats are the most suitable for this network traffic. The amount of darknet packets received are based on the placement of the sensors, the size of the monitored IPs and the configuration setup. In terms of placement, previous studies has shown that the traffic collected by two different but equally sized darknet is not the same. In terms of size of the darknet and network requirements, a study has shown that a small /24 sensor has approximately a rate of 9 packets per second, while a /16 sensor receives 75 packets per second, and a large /8 monitor receives over 5000 packets per second [16]. Second, in terms of deployment techniques, there are mainly three major darknet deployment approaches: i) the first is to simply send Address Resolution Protocol (ARP) replies for each dark address to the router. This technique, although it is simple, it works well when the darknet addresses are well known and are limited in size. When the unused addresses reach thousands or million, the monitoring activity becomes less efficient; ii) the second approach is considered more scalable and link a static range of IP address block to the monitor. This is a simple approach but it needs the dark address block to be specifically separated for analysis; iii) the third approach is done by forwarding all non-configured packets to the sensor. It is like forwarding all unused packets of an organization network to the sensor. The above mentioned methodologies deal with unused and reachable dark addresses. However, to capture unused and non-reachable packets, RFC 1918 [17] specifies some IP ranges that fit in this category. We refer the reader to [16] for more information regarding these techniques and how to implement them.

### 3.1.2 Darknet Data

In terms of protocols, TCP packets are the most common traffic found on darknet, UDP packets are the second and ICMP packets are the third [18, 19]. Moreover, in terms of threat types, the majority of the inferred traffic consists of scanning and worms activities [18, 19], backscatter traffic from DDoS victims [13] whereas the minority consists of misconfiguration of network devices such as routers. Concerning scanning and worm scenarios, the attack aims to scan or probe the Internet address space looking for gathering information (e.g., vulnerabilities in services) or infecting machines (e.g., downloading a malware), etc.. In regard to backscatter traffic scenario, the adversary sends a flood of requests signed by the IP address (spoofed) to the victim. Subsequently, the victim replies to the flood by sending backscattered (reply) packets to the sensors. In this scenario, the assumption is that the attacker chooses a spoofed IP address that belongs to the monitored darknet space. In another DDoS cases, the attacker flood the Internet with requests spoofed by the address of the victim. Hence, all open services (e.g., DNS resolver) found online will reply to the victim with amplified replies [36]. For all aforementioned cases, while generating these malicious activities, some of this traffic might reach the darknet sensor and hence become available for IT investigators. Subsequently, security operators can study the traffic (e.g., analyze the speed and impact of the attack) and understand the malicious strategy that are being used on the Internet. This exercise might help in detecting, preventing, mitigating and even attributing cyber attacks.

### 3.1.3 Darknet: Pros and Cons

Since darknet monitors solely unused address space, the major advantage of this approach is the security aspect on the monitor side. In other word, only uni-directional traffic (towards darknet sensor) is captured on the darknet. Recall that, unused address space do not communicate, and hence darknet is called passive monitoring. The security advantage of this approach is that darknet sensors are safe and not threaten by the attack. Further, darknet deployment is considered easy to deploy since it is passive (run in a listening mode) and do not require interaction with the adversary. On the other hand, the major disadvantage of darknet is the collection of data. Since there is no interaction with the adversary, investigators will not detect the complete attack scenario but only the first stage of it. For example, IT analysts can infer the scanning attempt for vulnerability but cannot save the malware (e.g., the executable) used while using this scanning attack. Therefore, compared to other monitoring systems, information gathered on darknet is considered low.

## 3.2 IP Gray Space

These addresses refer to devices that are not assigned to any host throughout a given time period (e.g. 1 hour, 1 day). In concept, IP gray space is very similar to darknet. The only difference is that IP gray space addresses are unused for a limited time only, whereas darknet addresses are permanently unused. Unlike darknet analysis techniques which can potentially be evaded, IP gray space might be harder to detect by an attacker since they are active and operating as a regular machine at some periods of time. The aim of this experiment is to confuse the adversaries while they are trying to gain some information from the targeted address space. For instance, after a successful compromising a machine that is running in active mode, the adversaries try to control the victim using a C&C. However, since this active machine might switch to inactive mode, the security operators might trick the attackers (e.g., botmaster) and infer their malicious next-step activities. Obviously, the assumption is that adversaries do still believe that the compromised machine is still under their control [20, 21].

### 3.2.1 IP Gray Space Deployment

As mentioned above, IP gray space addresses have mainly two status, namely, active and inactive. The analysis of IP gray space is done when the addresses are in passive (inactive mode). Therefore, in order to deploy IP gray space, you need a setup similar to darknet deployment in Section 3.1.1 when the addresses are in the inactive mode. However, when the addresses switches to active mode, the setup is switch to a regular active hosts online. Another way to deploy IP gray space, is simply disconnect an active machine for a limited period of time and sniff all incoming packets destined to its addresses. In literature, there is no detailed guideline to follow on deploying IP gray space network. For further information on IP gray space, we refer the reader to [20].

### 3.2.2 IP Gray Space Data

Since IP gray space is darknet for a specific period of time or has both active and inactive modes, collected inactive traffic can be darknet data (DDoS, scanning, misconfiguration) plus active traffic (Botnet C&C, malware) data. Therefore, assuming the IP gray space deployment is correctly done, analyzing its space can be more relevant and insightful than examining darknet data. For example, since darknet monitor only unused IP addresses (in passive/inactive mode), it is unlikely to find on its sensors interactive activities such as botnet/malware communication. In contrary, IP gray space monitoring can infer these communication as its addresses can run in active mode for a certain period of time.

### 3.2.3 IP Gray Space: Pros and Cons

Obviously, IP gray space monitoring holds the disadvantages and advantages of darknet while running in passive mode. However, while running in active mode, the IP gray space might have the following pros and cons. Regarding the advantages, as mentioned before, more data will be collected such as malware/botnet communication and hence more insights can be inferred. Regarding the disadvantages, running in active mode will provide adversaries with capabilities of attacking monitoring servers (sensors). Therefore, the deployment of IP gray space requires more attention and has more challenges than deploying darknet.

## 3.3 Honeypot

This is a computer system, mostly connected to the Internet that is configured to trap attackers. Honeypots are similar in nature to darknet with more specific goals. Honeypots, in general, require more resources than darknet since the aim is to interact with the adversary. There are 3 major types of honeypots, namely low, medium and high interactive honeypots. Types are differentiated based on their interactivity level with the initiator of the communication. On one hand, a low-interactive honeypot, is a simple solution configured to interact with the intruder at a basic level by emulating services (e.g., send *ECHO Reply* message to an *ECHO Ping Request*). Further, a medium-interactive honeypot is similar to low-interactive one but with further interactions and more emulated services for more data capturing and analysis (e.g., reply *SYN ACK* to a *SYN* request). On the other hand, a high interactive honeypot is a computer system that do not emulate some services. Instead, it runs a complete vulnerable or non-patched operating system and applications such as an OS version on a virtual machine. Note that a collection of honeypots form a honeynet [22].

### 3.3.1 Honeypot Deployment

Similar to other monitoring systems, honeypot deployment depends on several factors such as the location of the sensor, configuration of the sensor, and most importantly, the type of the sensor (low, medium, high). First, a major factor in honeypot deployment is choosing the location. A good practice exercise recommends installing several honeypot in separate locations and separate from the production system to prevent liability issues [23]. The more distributed are the sensors, the better are the extracted insights and the vision. Second, configuring a honeypot changes based on the need of the analysts. For instance, setting up a highly-interactive honeypot with capabilities to detect botnet is less challenging than deploying a low-interactive honeypot to monitor solely scanning activities. In any case, the deployment of low-interactive honeypot is close to darknet deployment whereas deploying high-interactive honeypot must be done in a way to emulate the operation of a regular machine. Therefore, a practical way to deploy high-interactive honeypot is to run the trap on a virtual machine (VM) [24] and run this service in a safe environment. A practical honeypot deployment is found in [25].

### 3.3.2 Honeypot Data

Honeypot is probably the best source for collecting security information due to the fact that these sensors can save the complete (bi-directional) communication or session with the adversary. Therefore, through honeypot, traces can track all attack stages such as scanning/probing for vulnerabilities, exploits, P2P and C&C communications, malware download activities, storing executable malicious codes, drop locations, etc. Several researchers investigate honeypot and extract various useful information on adversaries operation and strategies. For more information on honeypot data collection we refer the reader to [22, 26].

### 3.3.3 Honeypot: Pros and Cons

Honeypots, similar to any other trap-based monitoring system, have their own pros and cons. In terms of advantages, despite the fact that honeypot deployment is more complex than passive monitoring systems such as darknet, however, the technology today made implementing and running a honeypot a simple and flexible monitoring system. Most importantly, honeypots capture various security information regarding adversaries methodologies and techniques. The honeypot stored information is larger and more insightful than the darknet ones. In terms of the disadvantages, honeypots, since they interact with adversaries, this make this operation challenging. For instance, security operators must be very careful to avoid being detected by the hacker and/or get infected or/and become compromised. If the attackers discover any suspicious monitoring services, they might block the traced communication or even send irrelevant information to confuse investigators. Therefore, deploying the honeypot sensor might require more attention than other trap-based sensors such as darknet where monitoring do not require interaction with the adversary [24].

### 3.4 Greynet

This network is sparsely populated with darknet inactive addresses interspersed with active IP addresses. In other word, greynet uses both darknet (passive) and honeypots (active) on the same monitoring space during the same period of time. The purpose is to make the monitored IP block more attractive trap for the attacker. For example, a range of IPs that have both darknet and other active sensors running some fake services (i.e., reply to *ICMP ECHO* requests) might trick the attackers and have them think that the whole range of IPs in this monitored block is an appropriate organization target.

### 3.4.1 Greynet Deployment

Obviously, since greynet is an address space containing both darknet and honeypot, this means that deploying a greynet require deploying both darknet and honeypot. Therefore, deploying greynet is considered the most complex among all aforementioned trap-based monitoring systems. For instance, security analyst must distinguish between the passive and active monitoring traces and check for a possible link among them; at first, attackers might probe unused addresses (passive host) looking for vulnerable services and at second, the same or even other attackers might target a honeypot (active host) while searching for an active bot. This scenario might confuse investigators while collecting and interpreting this case to infer the pattern and the link among traces, if any. Research in this area is still required, for more information on greynet, we refer the reader to [27].

### 3.4.2 Greynet Data

Through Greynet, researchers and security operators can benefit from the fact that darknet and honeypot traces can be trapped. Therefore, in terms of passive monitoring, greynet can detect scanning, DDoS activities as well as misconfiguration. Moreover, in terms of active monitoring, greynet can identify botnet C&C activities, malware communication, drop locations of stolen information, etc. In a nutshell, greynet can trap various malicious activities and strategies that covers from the first until the last action of a malicious activity.

### 3.4.3 Greynet: Pros and Cons

Clearly, greynet holds the pros and cons of both darknet and honeypot monitoring systems. In regard to the advantage of greynet; first, a greynet is probably the most attractive space for adversaries to the fact that it represents a typical organization network having both active and inactive hosts. Hence, the more data is available for analysis. More importantly, greynet can be used to identify various scanning techniques and strategies [21]. On the other hand, some of the disadvantages behind greynet analysis is the amount and the variety of traces available for analysis. Consequently, analyzing greynet traffic is similar to monitoring a network organization data. The analysis of greynet, in contrary to darknet for instance, is complex to the fact that the traffic is coming from various types of sensors and hence mixing various types of attacks and strategies.

### 3.5 Honeytokens

Honeytokens were introduce in 2003 by Augusto Paes de Barros. Honeytokens are technologies that go beyond regular honeypots. In fact, honeytokens are honeypots without machines (sensors, computers, etc.). In other word, honeytokens are digital entities. Therefore, honeytokens have all the advantages of traditional honeypots and extend their capabilities beyond physical machines [29]. In practice, honeytokens can be, but not limited to, the following digital entities: a credit card number, a bogus login or any other information that can be stolen. The concept falls behind the following steps: First, security operators publish or make the aforementioned information accessible to adversaries. Second, assuming that hackers take control over these information and hence, security operators can monitor, track and trace-back hackers' activities.

### 3.5.1 Honeytokens Deployment
Since honeytokens deployment do not require physical entity, running them is considered simple and fast. In order to elaborate more on the deployment of honeytokens, we provide the following scenario. Imagine cyber security experts are investigating online credit card thefts. Therefore, for monitoring such malicious activities, the experts provide a bogus credit card information online (e.g., on a specific bank database) and wait for malicious activities on this bogus data. Any activities that run on this credit card information is considered a violation to the system's usage policy and hence needs to be investigated. It is noteworthy to note that in general, there is no designed algorithm or configured rules to deploy a honeytoken [29].

### 3.5.2 Honeytokens Data
Honeytokens, similar to any other monitoring system, do not solve all security issues. However, it can help in some security applications such as identifying source of attack and tracing the motive and the behavior of adversaries. Therefore, honeytokens analyzed data contains mainly traces of the attack, namely, handling stolen information, storing stolen information, tracing-back adversaries and their networks, discovering malicious users or users with malicious intent, etc. It is significant to note that honeytokens traps can be in the form of any file or information such as Office documents (PowerPoint, Excel, Word), pdf, databases, SIN, credit card info, bank accounts, signatures, login credentials, etc.

### 3.5.3 Honeytokens: Pros and Cons
With regard to the advantage of honeytokens, as mentioned before, this technology is considered easy to deploy and extremely effective with low cost. Honeytokens accuracy in detecting malicious users is higher than other trap-based monitoring system to the fact that its digital trap is more attractive and realistic to adversaries since it provide them assets (e.g., financial). For instance, from a law enforcement point of view, tracing back a bogus credit card might have more impact than inferring scanning activities. Further, a major advantage of this technology is the deployment which do not require configuration or/and setup of hardware devices. Hence, this technology operates dynamically and in a mobile manner. Instead of deploying a sensor on a specific IP address space, honeytokens operate independent of their locations (e.g., web-based). In regard to the disadvantage of this technology, honeytokens cannot trap all network-based information that destined to an IP address space due to the fact that, as mentioned earlier, honeytokens is a digital entity. Therefore, honeytokens fail to trap data regarding scanning activities, worm propagation, etc.

## 3.6 Comparative Study
In this section, we provide a comparative study of the trap-based monitoring systems, namely, darknet, IP grey space, honeypot, greynet and honeytokens.

### 3.6.1 Feature Comparison
This section compares the aforementioned monitoring systems based on the following features:

**Interactivity:** The interactivity is the measure of the interaction level between an adversary and the monitoring system. For example, a darknet, which is inactive and passive monitoring, do not require interaction with adversaries whereas a high-interactive honeypot require a high interactive communication.

**Complexity:** The complexity is the measure of the difficulties in setting up a monitoring system. In other word, how complex is to deploy a monitoring sensor. For instance, setting up a greynet which can run also honeypot, might be more complex than deploying solely a darknet.

**Data Collection:** The data collection measure the quantity of data collected from the trap sensor. For instance, obviously, high interactive honeypot which stores bi-directional communication might collect more information than a darknet that saves solely unidirectional data.

**Security:** The security feature measure the security level of implementing sensor on the monitoring side. For instance, the security of deploying a high interactive honeypot is high as the monitoring sensor might be threatened to become compromised.

| Monitoring System | Type | Interactivity | Complexity | Data Collection | Security |
|---|---|---|---|---|---|
| Darknet | IP-based | N | L | L | S |
| IP Gray Space | IP-based | N | L | L | S |
| Low-interactive Honepot | IP-based | L | L | L | V |
| Medium-interactive Honepot | IP-based | M | M | M | V |
| High-interactive Honepot | IP-based | H | H | H | V |
| Greynet | IP-based | N-LMH | LMH | LMH | S-V |
| Honeytoken | Digital-based | N-L | L | LMH | S |

N: Null – L: Low – M: Medium – H: High – S: Safe – V: Vulnerable

Table 1. Monitoring Systems - Feature Comparison

Table 1 provides a comparison of network monitoring systems based on several features, namely, type of sensor, interactivity with the adversary, complexity of deployment, data collection or information gathering and the security on the monitoring sensor. First, Darknet and IP gray space share similar features due to the fact that the latter is darknet but running for a specific period of time. These two safe monitoring systems are running in passive mode (null interactivity) and their deployment and data gathering entities are considered low compared to other monitoring systems. Second, in regard to honeypots, the interactivity, the complexity and the data gathering are mostly proportion to each other. For instance, the more the interaction with the adversary, the more complex is the design and the more data is collected. However, concerning the security aspect, all honeypots which have interactive feature can be vulnerable in terms of security. Third, since greynet consists of darknet and honeypot, it is considered the more comprehensive monitoring system and therefore, it could have all the possibilities in terms of interactivity, complexity, data collection and security. Finally, honeytokens are the only digital-based entity which has null to low interactivity with the adversary depending on the monitored digital entity and low complexity due to the simplicity of running it. Further, this secure and safe monitoring system can have low to high data collection depending on the investigation focus.

Although some trap-based monitoring systems could have hybrid capabilities (e.g., running in both darknet and honeypot), Table 1 provided a high level understanding on typical trap-based sensors and highlighted the thin and gloomy lines that separate among them.

### 3.6.1 Address Space Distribution
In this section we provide a comparison between IP-based monitoring systems. Therefore, we omit honeytokens monitoring sensors which are digital-based entities. In order to visualize the address space distribution, Figure 2 depicts the address space scenarios of darknet, IP gray space, honeypot and greynet.

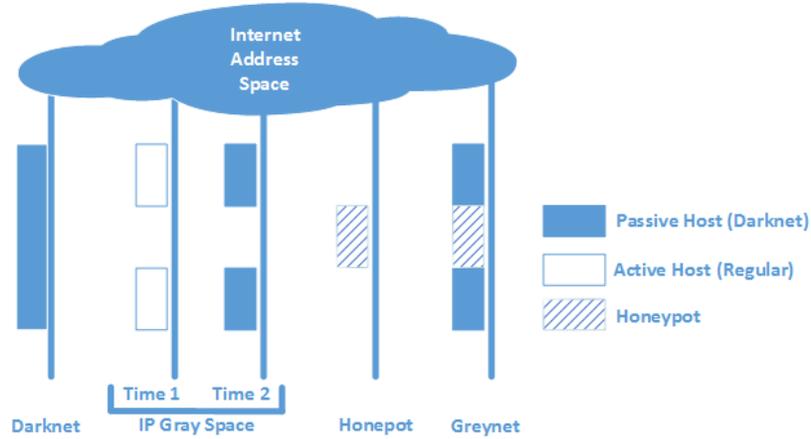
Figure 2. Monitoring Systems – Address Space Distribution

First, the darknet IP address space contains only unused addresses running in passive (inactive) mode. Second, the IP gray space (at time 2) is similar to darknet, however, the same address space was already active in a previous period of time (time 1). Third, honeypots can run in any types of network, either solely on a network as shown in Figure 2 or with active hosts or passive host. The latter case represents the final scenario which is the greynet address distribution.

## 3.7 Case Studies

In this section, we provide two case studies based on the aforementioned monitoring systems. The first is based on a darknet traffic analysis whereas the second focused on honeypots.

### 3.7.1 Darknet-Based Case Study: DAEDALUS-VIZ

The first case study represents DAEDALUS-VIZ [30] which is one of the large-scale cyber security projects (Nicter) that is based on the monitoring and the analysis of one type of trap-based monitoring system, namely, darknet. DAEDALUS-VIZ allows security operators to understand visually and in near real time the worldwide overview of security alerts. Moreover, the project provides high interactive and flexible tools to facilitate the investigation of large-scale cyber threats. An overview of the 3D visualization capability of this project is shown in Figure 3.

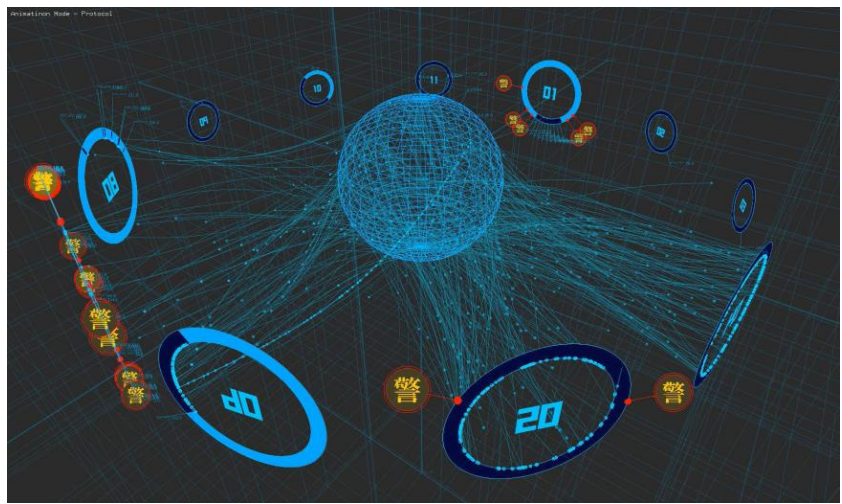
Figure 3. DAEDALUS-VIZ Overview [30]

DAEDALUS-VIZ visualizes various darknet threat, namely, internal darknet alerts that are based on local scans in addition to external darknet alerts that are based on global scans and backscattered traffic from victims of DDoS [13]. In addition, the project provides a novel technique to map IPv4 addresses on a sphere and other visualization shapes. Further, to the near real time processing, storing and displaying of information, DEADALUS-VIZ provides an overview system accompanied with drill down capabilities to investigate traces packet per packet. This is all running in an interactive platform which provides flexibility in viewpoint change, zooming capabilities, pausing/resuming services, among others. Most importantly, the system is customizable based on the investigator's need. Hence, security operators can change color, shape, size and position of the platform through a user friendly interface. Further, this visualization tool allows investigators to fine-grained data filtering operations.

### 3.7.2 Honeypot-Based Case Study: Situational Awareness of Large-Scale Botnet Probing Events

In this case study, we elaborate on a methodology to infer Botnet probing activities through Honeynets (multi honeypots). This case study do not solely infer the malicious activities, but also detect the strategy of the attack, its intention, its uniformity and coordination features. Figure 4 depicts the architecture overview of the botnet probing detection and inference system.

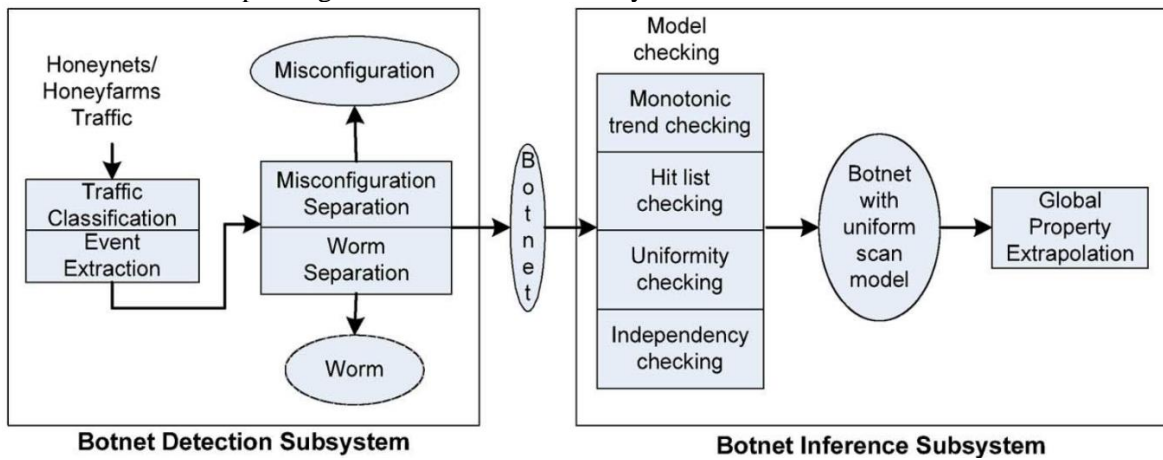

Figure 4. System Architecture [31]

As shown above, the architecture is mainly composed of two main elements, namely the detection and the inference subsystems. In the detection phase, the system takes honeypots information as input, than it classifies traffic before extracting events. The output of the later exercise divide the output into 3 categories, namely, misconfiguration, worm, and botnet. Since this system aims to infer botnet information, the second subsystem takes as input the botnet data for further analysis. The later subsystem is capable, through statistical and probabilistic algorithms, to infer the monotonicity of the botnet, the hit list data, in addition to the uniformity and independency of the strategy [31]. At the final stage, botnet traces that share similar properties are grouped as one botnet probing event. The output of the approach detected and inferred 203 botnet events that belong to 13 types of probes tackling various services such as HTTP, VNC, NetBIOS, MySQL, Telnet, etc.

## 4. SECURITY POLICIES AND LEGAL ISSUES

A cybercrime happens when both the action and the will are present. This section provides information on the security policies and legal issues while monitoring the Internet. Therefore, we aim to answer the following questions: Who has the capability to monitor the cyberspace? What are the privacy and security policies behind security analysis and deployment?

Since monitoring sensors (e.g., honeypot) are used to analyze data that is probably based on interactions with users (e.g., adversaries), these monitors might turn to legal responsibilities and several factors might turn them into liability. In this section, we provide an overview on the legal and security aspects of deploying, monitoring and analyzing monitoring data.

In the United States, there are at least three legal concerns to consider while deploying an interactive monitoring system. First, understanding the laws that restrict the owner of the security monitoring systems to monitor user activities. Second, addressing the risk and understanding that the monitoring system might be used by adversaries to harm others. Third, considering the fact that adversaries can argue that the monitoring service was operating undercover and entrapped them. In order to make things work perfectly, the owner of a trap-based monitoring system is recommended to ask a lawyer before implementing an interactive cyber sensor. Further, it is noteworthy to mention that cyber policies and restrictions might change depending on the country where the sensor is being deployed or/and the data is being analyzed.

## 4.1 Users Monitoring

Running and operating a monitoring system might be a simple exercise. However, having the legal right to monitor network activities of users might be the issue. Many sources might have restrictions on monitoring the cyber space such as privacy and employment policies, terms-of-service agreement, state and provincial laws, etc. Any violation of these laws might lead to civil liability and even criminal sanctions.

It is noteworthy to mention that monitoring cyber activities today takes a big attraction as it produces an exclusive insights for security operators, organizations, law enforcement agencies, governments and even adversaries.

### 4.1.1 Fourth Amendment

The *Fourth Amendment* limits the authority of government representatives to search or seize for evidence without first obtaining a search warrant from a judge. Since monitoring activities online could constitute a search and seizure, operating a monitoring system under a government agency direction might allow the *Fourth Amendment* of the US Constitution to restrict the monitoring process [22]. Further, if the *Fourth Amendment* law applies, the evidence obtained from monitoring systems can be suppressed at trial and even individual can face a lawsuit. Only individuals who have a reasonable expectation of privacy can criticize the aforementioned law and complain that the examination was unconstitutional under *Fourth Amendment*. In general, in a lawsuit, hackers do not have a reasonable motive to support their malicious actives. However, a private investigator, who is not under the government direction might analyze monitoring systems data without obeying to the above mentioned amendment.

### 4.1.2 Wiretap Act

The *Wiretap Act* forbids any individual to intercept or/and sniff network communications unless an exception registered in the act is applied. Even if an adversary do have a reasonable expectation of privacy while trapped in a monitoring system, under the *Wiretap Act*, this does not provide permission of monitoring. It is noteworthy to mention that violating this act can lead to civil liability and even a federal offence. Several exceptions belong to this act such as the *Provider Protection* and the *Consent of a Party* exceptions. Owners of monitoring systems could use these two exceptions to listen to user activities online. However, if the analysis is done under the government's authorities, the owners can utilize the *Computer Trespasser* exception to monitor cyber communication. *Provider Protection* exception allows the owner to monitor data if and only if the purpose of this monitoring activities is to protect the owner's rights or/and property from malicious use of the Internet such as fraud, theft, etc.

Note that, cyber security laws are still not robust due to the fact that this topic is considered new and technology is changing in a rapid manner. For instance, the laws do not include yet honeypot in the *Provider Protection* exception.

Concerning the *Consent of a Party* exception, the owner can monitor the traffic only after providing the consent from the other entity. By doing this act, the other entity agrees to be monitored. In practice, honeypot owners can post/display a consent banner for users who access their monitoring systems. In this case, all users who access this network will be able to see, read and accept the terms on the consent banner. In today's world, this is similar to the case when ISP customers call their company by phone and the voice operator inform the users that this phone call might be recorded for a certain reason. It is noteworthy to mention that consent banner might not work on all network services and must be specific or general depending on the service used [22].

### 4.1.3 Patriot Act
The *Computer Trespasser* is another exception that is part of the US *Patriot Act* that was passed in 2001. This exception allows government to monitor hackers in some cases without securing a warrant. This exception applies to users that are acting as a government agents to monitor adversaries' activities if: i) the network's operator has legally allow the capture; ii) the user who is performing the monitoring service is affiliated to a lawful investigation; and iii) if the person has a convincing reason and believe that the monitored traces will be appropriate to the investigation. This act might be useful when the honeypot is running under the operation of government or law enforcement agencies.

## 4.2 Harming others
While communication with hackers, monitoring system might be easily compromised. Therefore, security operators must be careful to reduce the risk of harming others unintentionally. Hackers can control monitoring systems, their bandwidth and network resources to turn them into malicious tools. Some good practices used on the monitoring systems include listening to their output traffic, controlling its parameters such as rates/bandwidth, filtering out suspicious activities and dropping blocking services/domains/IPs when necessary. This exercise require not just proper deployment of sensors, but also maintaining monitoring sensors' activities. Otherwise, monitoring systems can easily and rapidly be transformed to a network entity that belongs to adversaries (e.g., bot). Consequently, this entity can be used for storing malicious files, stolen information, and distributing malicious content such as child pornography and spam emails, etc.

## 4.3 Entrapment
The supreme court defines entrapment as "the conception and planning of an offense by an officer, and his procurement of its commission by one who would not have perpetrated it except the trickery, persuasion, or fraud of the officers" [32]. Entrapment can be a concern for honeypot owners due to the fact that criminals can use it in an attempt to evade conviction. However, a defendant that was not induced by law enforcement authorities to commit a crime cannot utilize the entrapment to defend the case. It is notable to mention that entrapment policies apply solely in the context of criminal prosecution and hence, it is not applicable to private security monitoring owners.

## 5. FUTURE RESEARCH DIRECTIONS
For future work, security operators are required to look deeply into, but not limited to, the following areas:

- **IPv6 Security:** IPv6 is going to be the new era of Internet address space due to the fact that IPv4 addresses are about to get congested. Further, hackers are always one step beyond security operators and hence they will focus on new systems to abuse it. IPv6 is one of the most promising targets [34]. For instance, amplified DDoS attacks, which are based on the size of the packets, can be more harmful when using IPv6 (rather than IPv4) since IPv6 packet is larger in size. Moreover, IPv6 packets have built-in packet-level encryption and considered more secure than IPv4 traffic, hence, it might cause more challenges for network operators to investigate malicious behavior within its traffic. As of today, the majority of the current deployment, monitoring and analysis is based on IPv4 traffic. Very few works are done on IPv6 [33]. Therefore, security operators and researchers are recommended to focus on this area by setting up more IPv6 monitoring sensors to uncover cyber attacks that are embedded in IPv6 traffic.

- **Cloud Security:** Cloud is going to be the first service used on the cyber space. On the cloud, the number of services, users and data is enormous. Therefore, adversaries will consider the cloud as the best place to attack due to the fact that enormous architecture means large impact and easy to obfuscate. Several cloud security issues exist today [35] and very few security measures are deployed. All security operators must think in an enormous scale while installing security services on the cloud. This might require merging several security operators and systems to cope with the cloud security challenge such as a centralized large-scale cyber monitoring service using trap-based monitoring systems as well as firewalls and antivirus tools.

## 6. CONCLUSION

Internet users are growing intensely in numbers due to the advanced in computer applications today such as social media, cloud services, online shopping, among others. Consequently, information sharing in general and security in particular became a challenge for network operators, Internet service providers, organizations, law enforcement agencies and governments. This provocation is raised due to the lack of big data handling, storing, monitoring and analyses. In order to cope with this challenge, several cyber monitoring systems are used on different IP layers such as networks and applications. In this work, we have discussed that one of the best techniques to monitor the cyber space is to trap adversaries by installing monitoring systems such as darknet, IP gray space, honeypot, greynet and honeytokens. In general, these systems run on unused but routable address space, and hence all traffic destined to them might be suspicious. Darknet and IP gray space operate in passive (inactive) mode whereas honeypot and greynet can operative in active mode. Among all the aforementioned trap-based monitoring systems, only honeytokens is based on a digital entity, whereas the others are focused on IP services. It is noteworthy to mention that some monitoring systems provide hybrid services and can operate in different modes (passive and active) like honeyd [28]. Since each type of monitoring system has its own advantages and disadvantages, a good practice is to deploy several services to cover all security aspects. Note that, these systems do not replace other security monitoring and detection services such as intrusion detection systems. In contrary, trap-based monitoring services must run in parallel with IDS and other security services such as firewalls and antivirus tools. Therefore, hybrid monitoring systems are considered to be the best options to choose while deploying a security monitoring system. This chapter discussed various cyber security threats that can be inferred by analyzing traces from the above mentioned monitoring systems such as scanning/probing activities, DDoS attacks, worm propagation, botnet C&C, and malware infections. In general, passive monitoring is decent in inferring scanning/probing and DDoS activities whereas active monitoring, which is interactive with the adversary, is more attractive to botnet and malware activities.

In general, we have seen that deploying passive network monitoring systems is easier than installing and configuring an active monitoring systems. Nevertheless, the easiest trap-based monitoring system to deploy is the honeytoken which is based on providing a bogus digital information such as fake credit card information to an adversary and then monitoring its activities to infer insights on the malicious behavior such as the intention of the attack. However, in general, implementing a trap-based monitoring system is becoming an easy task today as some commercial organizations provide services to install a ready-to-go sensor or even provide the extracted data extracted from the sensor even without deploying it at the client side. Finally, concerning the security policies and legal issues, we have seen that cyber security laws are still in a gray area due to the fact that cyber security is still considered a new topic and therefore several laws are not applicable to specific areas such as trap-based monitoring systems. As a result of the rapid and recent growth in cyber technology, more laws must be updates and new ones must be created.

# ADDITIONAL READING SECTION

- Krawetz, N. (2004). Anti-honeypot technology. Security & Privacy, IEEE, 2(1), 76-79.
- Pouget, F., Dacier, M., & Pham, V. H. (2005, March). Vh: Leurre. com: on the advantages of deploying a large scale distributed honeypot platform. In In: ECCE 2005, E-Crime and Computer Conference.
- Pouget, F., & Dacier, M. (2004, May). Honeypot-based forensics. In AusCERT Asia Pacific Information Technology Security Conference.
- Joshi, R. C., & Sardana, A. (2011). Honeypots: A New Paradigm to Information Security. Science Publishers.
- Curran, K., Morrissey, C., Fagan, C., Murphy, C., O'Donnell, B., Fitzpatrick, G., & Condit, S. (2005). Monitoring hacker activity with a Honeynet. International Journal of Network Management, 15(2), 123-134.
- Yegneswaran, V., Barford, P., & Paxson, V. (2005). Using honeynets for internet situational awareness. In Proc. of ACM Hotnets IV.

# KEY TERMS & DEFINITIONS (SUBHEAD 1 STYLE)

**DDoS:** A Distributed DoS (DDoS) is among the most sever cyber attack where adversary bombards the victim with large amount of data/requests in order to exhaust its bandwidth or/and network and host resources.

**DNS amplification:** One type of DDoS attack where the attacker try to send amplified replies to the victim in order to exhaust its bandwidth or/and network and host resources.

**Malware:** A malware is a piece of code that is malicious.

**ISP:** An Internet Service Provider is a certified organization for providing Internet services.

**Firewall:** A firewall is a computer hardware or software that has an aim to control the security of the network by applying rule set on the incoming and outgoing traffic.

**Antivirus:** An Antivirus is a computer software with an aim to prevent, detect, identify and remove malicious malware and viruses.

**SNMP:** Simple Network Management (SNMP) protocol is used for managing devises in a network.

**IRC:** Internet Relay Chat (IRC) is a computer program used to transfer text message between two entities.

**P2P:** Pear-to-Pear (P2P) is a communication technology where each client can be either client or server.

**CPU:** The Central Processing Unit (CPU) is a hardware device inside a computer machine that is responsible on performing basic and logical computer operations.

**Root:** In Unix Operating Systems, Root is equivalent to a super user with usually full administrative privileges.

**Backscatter:** In a communication protocol, backscatter traffic are packets or traffic that is based on the reply packets (e.g., SYN-ACK, ACK).

**Spoof:** IP address spoofing is the phenomenon of forging the source IP address for concealing the identity of the sender or impersonating another identity (e.g. victim).

**ARP:** In multiple access networks, the Address Resolution Protocol (ARP) is designed to resolve network layer addresses to link layer addresses.

**TCP:** The Transmission Control Protocol (TCP) is one of the major transport protocol that operates on top of the IP protocol. The aim of this protocol is to transmit data in a reliable manner.

**UDP:** Similar to the TCP protocol, the User Datagram Protocol (UDP) is another core element on the transport layer. The aim of this protocol is to transmit data in a fast manner.

**ICMP ECHO:** The Internet Control Message Protocol (ICMP) is a core element on the IP layer. This protocol is responsible on the control and error handling of messages among devices such as routers. ICMP messages are many. ECHO message is one type of ICMP.

**VM:** A Virtual Machine (VM) is a computer program that emulates a real (physical) machine. A virtual machine can be used for several purposes such as (testing, learning, etc.).

**HTTP:** The Hypertext Transfer Protocol (HTTP) is a fundamental data communication process for the World Wide Web (www). It consists of an application protocol for distributing and collaborating hypermedia data systems.

**VNC:** A Virtual Network Computing is a remote desktop sharing protocol used to control another device.

**NetBIOS:** A Network Basic Input/output System (NetBIOS) provides communication services related to session layer of the Internet and permitting Local Area Network communications among devices.

**MySQL:** MySQL is one of the most famous open source database with an aim to store, organize and retrieve information.

**Telnet:** Telnet is a network service that provides an interactive text-based communication facilities through a virtual terminal network.